\documentclass{svproc}
\usepackage{graphicx}%
\usepackage{multirow}%
\usepackage{amsmath,amssymb,amsfonts}%
\usepackage{mathrsfs}%
\usepackage[title]{appendix}%
\usepackage{xcolor}%
\usepackage{textcomp}%
\usepackage{manyfoot}%
\usepackage{booktabs}%
\usepackage{algorithm}%
\usepackage{algorithmicx}%
\usepackage{algpseudocode}%
\usepackage{listings}%
\usepackage{subcaption}

\usepackage{url}

\def\orcidID#1{\unskip$^{[#1]}$} % added MR 2018-03-10

\begin{document}
\mainmatter              % start of a contribution
%
%\title{Human Mobility as a Complex Process: \\ A Process Mining Approach}
%\title{Structuring Urban Mobility with Object-Centric Process Mining}
\title{Revealing Process Structure \\ in Urban Mobility Networks}
\titlerunning{Human Mobility as a CP: A PM approach}  % abbreviated title (for running head)
\author{Khristina Filonchik\inst{1}\orcidID{0009-0004-8205-472X} \and Jose Pedro Pinto\inst{1} \and Flávio L. Pinheiro\inst{1} \and Fernando Bacao\inst{1}}
\authorrunning{Khristina Filonchik et al.} % abbreviated author list (for running head)

\institute{NOVA IMS –  NOVA Information Management School (NOVA IMS), Universidade NOVA de Lisboa, Campus de Campolide, 1070-312, Lisboa, Portugal\\
\email{20230692@novaims.unl.pt}}

\maketitle              % typeset the title of the contribution

\begin{abstract}
Urban mobility is a multi-entity system that involves travelers, transport modes, and infrastructure. Beyond conventional origin/destination analysis, this paper investigates how process mining can structure and interpret mobility behavior from event data. Using Call Detail Records (CDRs) from Oeiras in the Lisbon metropolitan area (Portugal), we construct both case-centric and object-centric event logs and discover models that summarize flows and typical durations. Results show that most trips are intra-municipal, while inter-municipal flows connect strongly to neighboring areas, with typical inter-parish travel times of about 20 minutes. The object-centric perspective explicitly links trips and transport modes, revealing mode-specific duration differences (e.g., bus vs. car) that inform multimodal planning. Our contributions are: (i) a reproducible pipeline to transform CDRs into process mining artifacts, (ii) empirical evidence that mobility data exhibit a process-like structure, and (iii) the added value of object-centric models for multimodal analysis. Limitations include the low spatial precision of CDRs (tower-sector level) and heuristic transport-mode labels. Future work will integrate transport-network context (e.g., stations and routes) and model object-centric logs as heterogeneous graphs to enable richer and more reliable analysis.

\keywords{human mobility, process mining, object-centric log, call detail records}
\end{abstract}

\section{Introduction}\label{sec1}

The study of human mobility has entered the Big Data era with the availability of location-enabled traces such as mobile networks, GPS, and smart cards ~\cite{mokbelMobilityDataScience2024,pappalardoFutureDirectionsHuman2023}. These sources provide large-scale evidence of how people move, but also introduce spatial and temporal biases. Traditional origin–destination (OD) surveys remain widely used to synthesize flows~\cite{andaTransportModellingAge2017}, but are sample-based and overlook sequential, multimodal, and contextual aspects of mobility. Effective planning therefore requires methods that not only observe where people travel, but also capture how and under what conditions.

Complex networks offer powerful abstractions of urban flows, from structural backbones to community detection~\cite{BARBOSA20181}. Recent work quantifies accessibility through network measures (e.g., the 15 minute city ~\cite{su15043772,abbiasovAuthorCorrection15minute2024}) or applies graph learning to visualize and forecast mobility ~\cite{huang2025daymultiscalespatialtemporaldecoupled,mitra2025graphnetworkmodelingtechniques}. These approaches emphasize structure, but often neglect the temporal logic of trips. 

Process Mining (PM) addresses this gap by modeling event data as sequences of cases, events, and timestamps~\cite{vanderAalst2022}. Although widely used in domains such as logistics, manufacturing, and healthcare, PM is rarely applied to urban mobility, where the traces are noisy, sparse, and highly variable. This paper demonstrates how PM can be applied to large-scale Call Detail Records (CDRs). We construct both case-centric and object-centric event logs, compare their ability to capture sequential and multimodal patterns, and discuss their complementarity to network-based approaches. Using a case study of Oeiras (Lisbon metropolitan area, Portugal), we highlight both opportunities and limitations, positioning PM as a step toward integrating process-oriented analysis into complex-systems research.

\section{Related Work}\label{sec2}
Graph-based methods model mobility as networks of locations and movements. The location and activity graphs show that people revisit limited sets of places and that transitions can be quantified with graph-theoretical metrics ~\cite{alessandrettiEvidenceConservedQuantity2018,longActivityGraphsSpatial2023,martinGraphbasedMobilityProfiling2023}. These studies emphasize structure, but do not capture the temporal unfolding of trips, which is essential for process-oriented analysis~\cite{barthélemySpatialNetworks2011}.

PM provides such a perspective by discovering sequential patterns from event logs ~\cite{vanderAalst2022}. Applications in mobility are limited, but growing: tram smart cards in Porto~\cite{ribeiroMultidimensionalSubgroupDiscovery2024}, bus system in Montevideo~\cite{delgadoProcessMiningImproving2023}, bike sharing in New York~\cite{drosouliProcessMiningApproach2020}, and commuting with GPS traces ~\cite{yousfiDiscoveringCommutePatterns2019}. These works show PM potential, but are case-specific and rely on case-centric logs. On scale, such logs produce highly variable “spaghetti” models ~\cite{diamantiniDiscoveringMobilityPatterns2017} and do not represent the multi-entity nature of mobility. Recent advances in Object-Centric PM (OCPM) ~\cite{vanderaalstMOREREALISTICSIMULATION2023,bertiOCPMAnalyzingObjectcentric2023} address this limitation by linking events to multiple entities. While OCPM has shown promise, it remains a new research stream with limited applications to large-scale, high-dimensional data.

\section{Dataset and Methodology}\label{sec3}

\subsection{Methodology}
%\label{sec:antenna-spreading}
The data used in this study were provided by a Portuguese telecommunications provider for research purposes. It consists of call and SMS events. Due to strict privacy regulations, all data pre-processing was performed collaboratively with the company on their internal servers using the open source Python packages infostop~\cite{aslak2020infostop}\footnote{\url{https://github.com/ulfaslak/infostop}} and trackintel~\cite{martinTrackintelOpensourcePython2023}\footnote{\url{https://github.com/mie-lab/trackintel}}. For this study, we focus our analysis on a subset of the full dataset, specifically selecting the trips that were started at the municipality of Oeiras within the Lisbon metropolitan area (Portugal) and covering the period from February to March 2024. With a population of $171,658$, Oeiras - like many municipalities outside the city center of Lisbon - offers a valuable case for studying inter-municipal mobility, particularly in terms of travel destinations and frequency of trips. 

\begin{figure}[h]
\centering
\includegraphics[width=12.5cm]{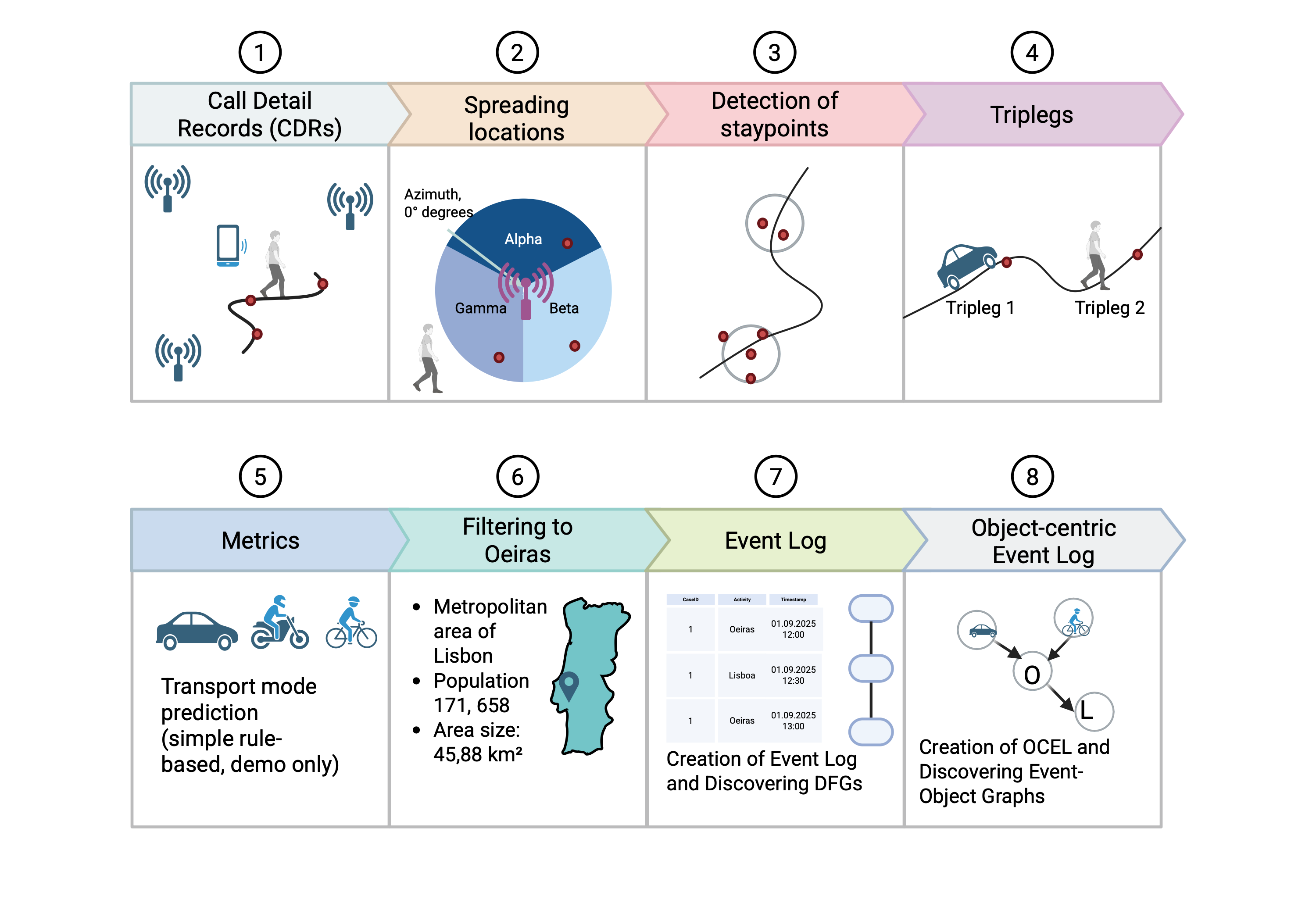}
\caption{Methodology~\cite{filonchik2025biorender}}
\label{fig:method}
\end{figure}

The raw dataset covers the Lisbon metropolitan area and several surrounding municipalities with $43,529,555$ events of $129,032$ users. Figure~\ref{fig:method}
summarizes the entire data transformation methodology used in this work. 

CDR data provide only the ID of the serving cell, not the exact location of the user. We approximate positions by sampling deterministic pseudo-locations within tower sectors defined by coordinates, azimuth, and radius, clipped to land areas along the Tagus river to avoid implausible placements~\cite{ellingson2025method}. To extract stays from sparse and noisy CDR traces, we applied infostop~\cite{aslak2020infostop} - flow-based network community detection algorithm, which detected $18,860,722$ staypoints (vs. $4,883,096$ with trackintel, $2,900,450$ with scikit-mobility\footnote{\url{https://github.com/scikit-mobility/scikit-mobility }}~\cite{pappalardoScikitmobilityPythonLibrary2022}).

Following the trackintel definition~\cite{zotero-item-6337}, once staypoints are detected with infostop, each staypoint is assigned a unique identifier denoting an individual stationary segment. Based on these staypoints, triplegs are derived as contiguous movement segments without a change in the mode of transport or vehicle~\cite{hensherHandbookTransportModelling2007}. Finally, triplegs form trips that can consist of multiple triplegs. The final dataset covers $41,025$ users and $329,013$ trips (February to March 2024), spatially linked to Lisbon's parishes and municipalities, with Oeiras as the focal origin. 

\subsection{Event log generation for PM}
A traditional event log or case-centric is typically defined as a multiset of traces \cite{vanderAalst2022}, where each \textbf{trace} represents a sequence of events related to a single case (e.g., a trip). 
\begin{definition}[Case-Centric Event Log]

Let $\mathcal{C}$ be a set of \emph{case identifiers} (e.g., trip ID), $\mathcal{A}$ a set of \emph{activity labels} (e.g., visited locations) and $T$ a set of \emph{timestamps}.  
An \textbf{event} is a tuple $e = (c, a, t) \in \mathcal{C} \times \mathcal{A} \times T$.  

A \textbf{trace} for case $c \in \mathcal{C}$ is a finite sequence $\sigma_c = \langle e_1, \ldots, e_n \rangle$ such that each $e_i = (c, a_i, t_i)$ and $t_i \leq t_{i+1}$ for all $1 \leq i < n$.

An \textbf{event log} is a multiset of such traces:
\[
L = \{ \sigma_c \mid c \in \mathcal{C} \}
\]
\end{definition}

\paragraph{Example:}
\[
\sigma = \langle (\text{trip\_1}, \text{Oeiras}, t_1), (\text{trip\_1}, \text{Lisbon}, t_2) \rangle
\]

The object-centric event log follows the formal OCEL specification \cite{bertiOCELObjectCentricEvent}\cite{vanderaalstObjectCentricProcessMining2019}, which was used to build a mobility-oriented OCEL for this study. An object-centric event log (OCEL) generalizes the traditional model by allowing each event to relate to multiple objects of different types (e.g., trips, transport modes, locations). 

\begin{definition}[Object-Centric Event Log]
An OCEL is a tuple\cite{bertiOCPMAnalyzingObjectcentric2023}
\[
L = (\mathcal{E}, \mathcal{O}, \mathcal{A}, T, \mathcal{OT}, \pi_{\mathrm{type}}, R)
\]
where:
\begin{itemize}
  \item $\mathcal{E}$: event identifiers, each $e = (\mathrm{id}_e, a, t) \in \mathcal{E} \times \mathcal{A} \times T$,
  \item $\mathcal{O}$: object identifiers, $\mathcal{OT}$: object types,
  \item $\pi_{\mathrm{type}} \colon \mathcal{O} \to \mathcal{OT}$: typing function,
  \item $R \subseteq \mathcal{E} \times \mathcal{O}$: event-object relation.
\end{itemize}
\end{definition}

\paragraph{Example:}
\[
\begin{aligned}
e_1 &= (\texttt{e1}, \text{"OEIRAS"}, t_1) \\
R(e_1) &= \{\texttt{Bus}\}, \\
\pi_{\mathrm{type}}(\texttt{trip\_1}) &= \texttt{Bus} 
\end{aligned}
\]

For event abstraction, we used high-level geographic units, municipalities, as event labels, allowing us to examine how different mobility modes are related to travel flows between regions.

\section{Experiments and results}
Experiments were conducted with the open-source package PM4Py~\cite{bertiPM4PyProcessMining2023}\footnote{\url{https://github.com/process-intelligence-solutions/pm4py/tree/release}}.
Table~\ref{tab:cdrs1} presents summary statistics for the case-centric and object-centric (OCEL) logs, including the number of cases / objects, events, and variants / event-object relations. Conformance was assessed through token-based replay, which simulates each trace in the discovered Petri net and counts missing/remaining tokens to compute fitness in [0,1] (1 = perfect alignment)~\cite{berti2021novel}.
In our data, the case-centric log achieves fitness = 1.00, i.e., all traces are perfectly reproduced by the model.
\begin{table}[h]
\caption{Log statistics}
\label{tab:cdrs1}
\centering
%\resizebox{\columnwidth}{!}{%
\begin{tabular}{c l l l l l} 
\toprule
Log type & NumCases/Objects & NumEvents & Var./ObjTyp. & Ev./Obj. rel.  \\
\midrule
Event log, municipality & $329,013$ & $642,440$ & 10 & - \\
Event log, parish & $329,013$ & $658,026$ & 815 & - \\
OCEL, municipality & $658,026$ & $329, 023$ & $15$ & $1,316,052$\\
\bottomrule
\end{tabular}
%}
\end{table}

\subsection{Directly Followed Graphs (DFGs)}
A Directly-Follows Graph (DFG) presents the most straightforward depiction of process models~\cite{vanderaalstProcessMiningHandbook2022}. In this type of graph, the nodes symbolize individual activities, while the arcs illustrate the connections among different activities.
We use DFGs to summarize how one mobility state follows another. 

\begin{figure}[h!]
\centering
\begin{subfigure}[t]{0.88\textwidth}
    \centering
    \includegraphics[width=0.7\linewidth]{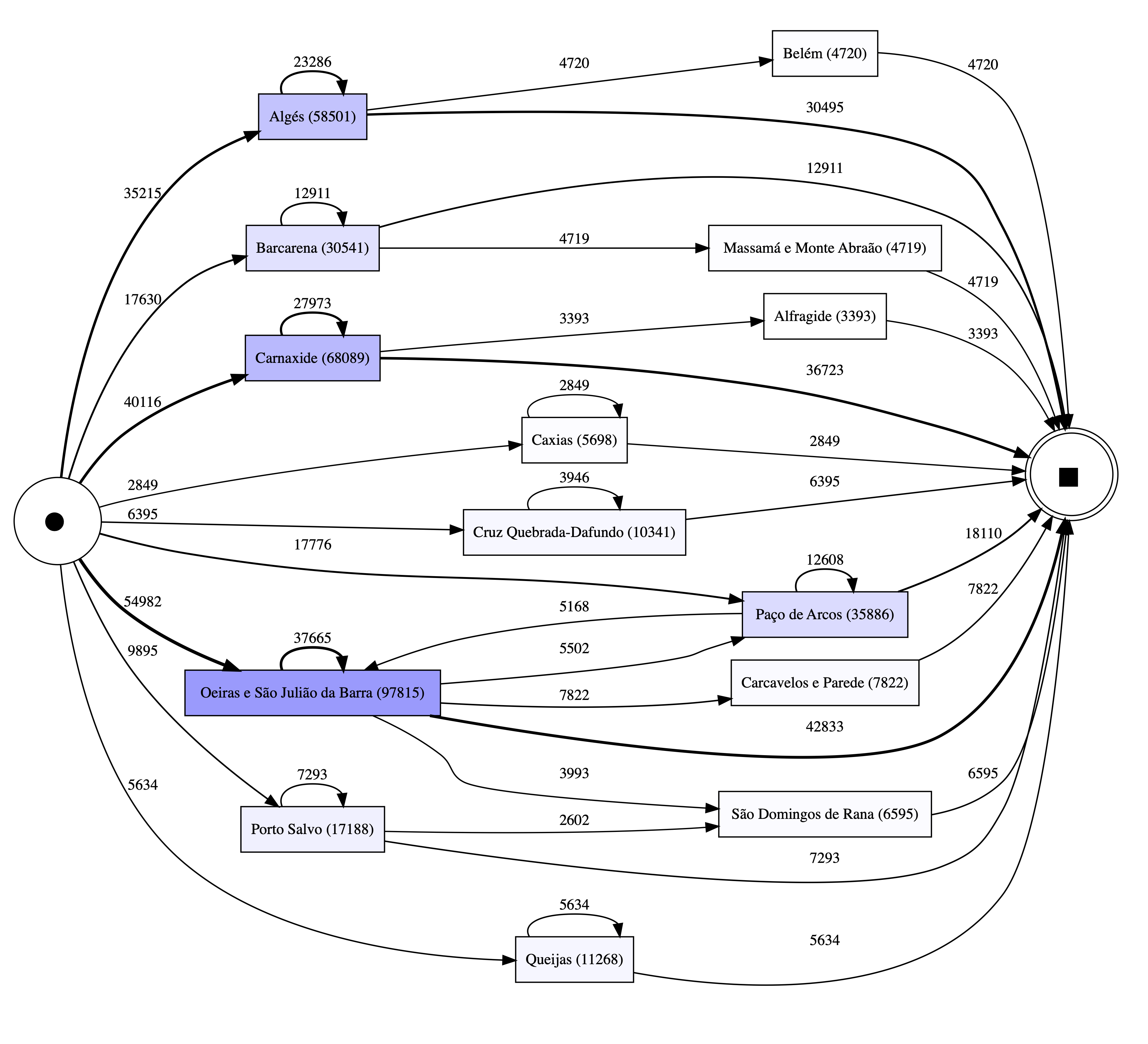}
    \caption{Top 20 variants by frequency}
    \label{fig:heu_top50fr}
\end{subfigure}
\hfill
\begin{subfigure}[t]{0.88\textwidth}
    \centering
    \includegraphics[width=0.7\linewidth]{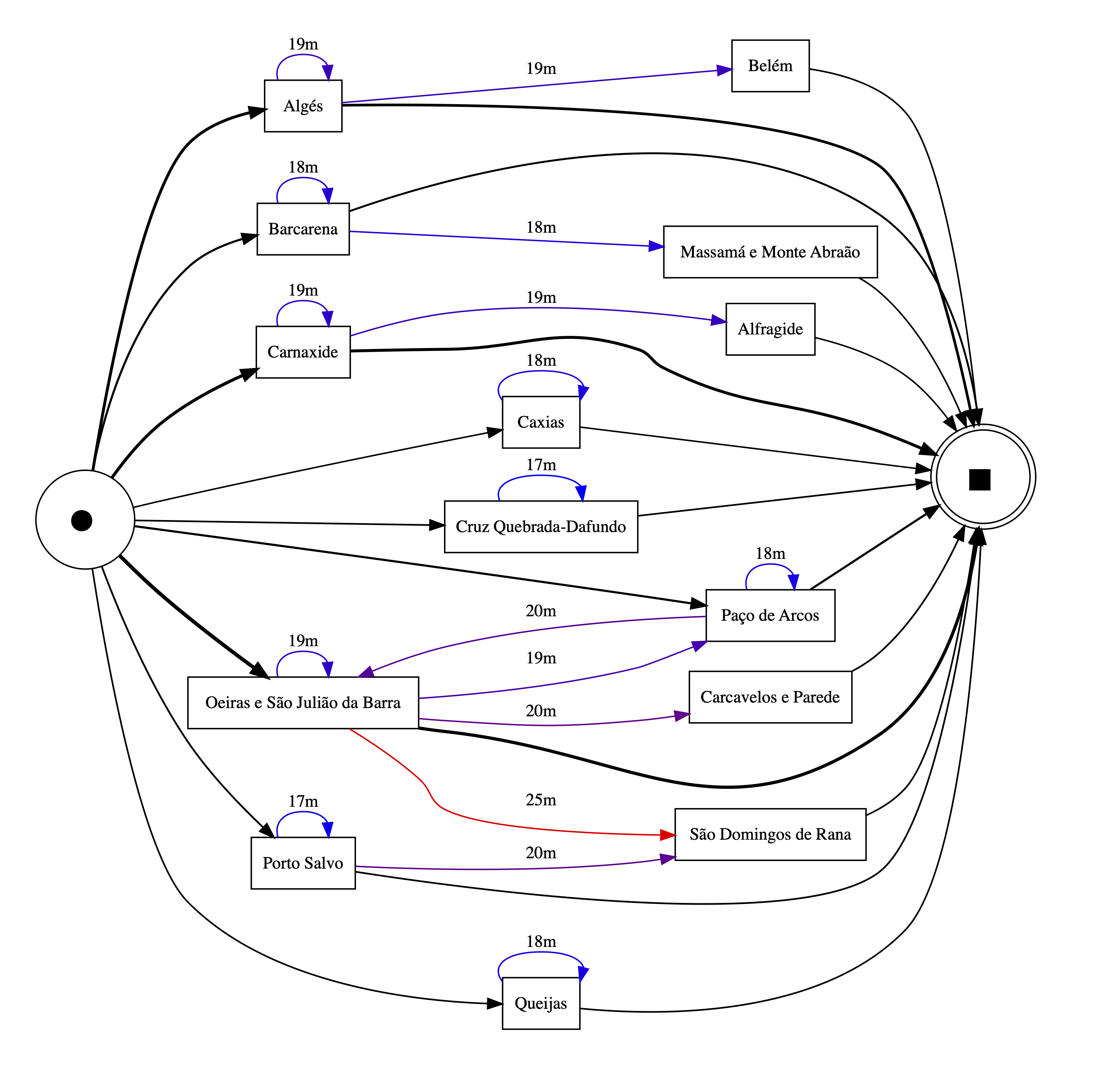}
    \caption{Top 20 variants with mean duration}
    \label{fig:heu_top50}
\end{subfigure}
\caption{Parish-level Directly-Follows Graphs: (a) Top 20 by frequency and (b) Top 20 with mean duration.}
\label{fig:heu_top20_combined}
\end{figure}

Figure~\ref{fig:heu_top50fr} shows the top-20 parish variants. The most common origins are Oeiras e São Julião da Barra (97,815), Carnaxide (68,089), Algés (58,501), Paço de Arcos (35,886), Barcarena (30,541), and Porto Salvo (14,586). Many trips connect neighboring parishes (e.g., Carcavelos–Paço de Arcos). A strong self-loop in Oeiras e São Julião da Barra indicates many within-parish trips; their average duration is 19 minutes (Fig.~\ref{fig:heu_top50}). In general, many intra-Oeiras and near-neighbor trips (e.g., from Algés) take about 20 minutes.

\begin{figure}[h!]
\centering
\begin{subfigure}[t]{0.49\textwidth}
    \centering
    \includegraphics[width=\linewidth]{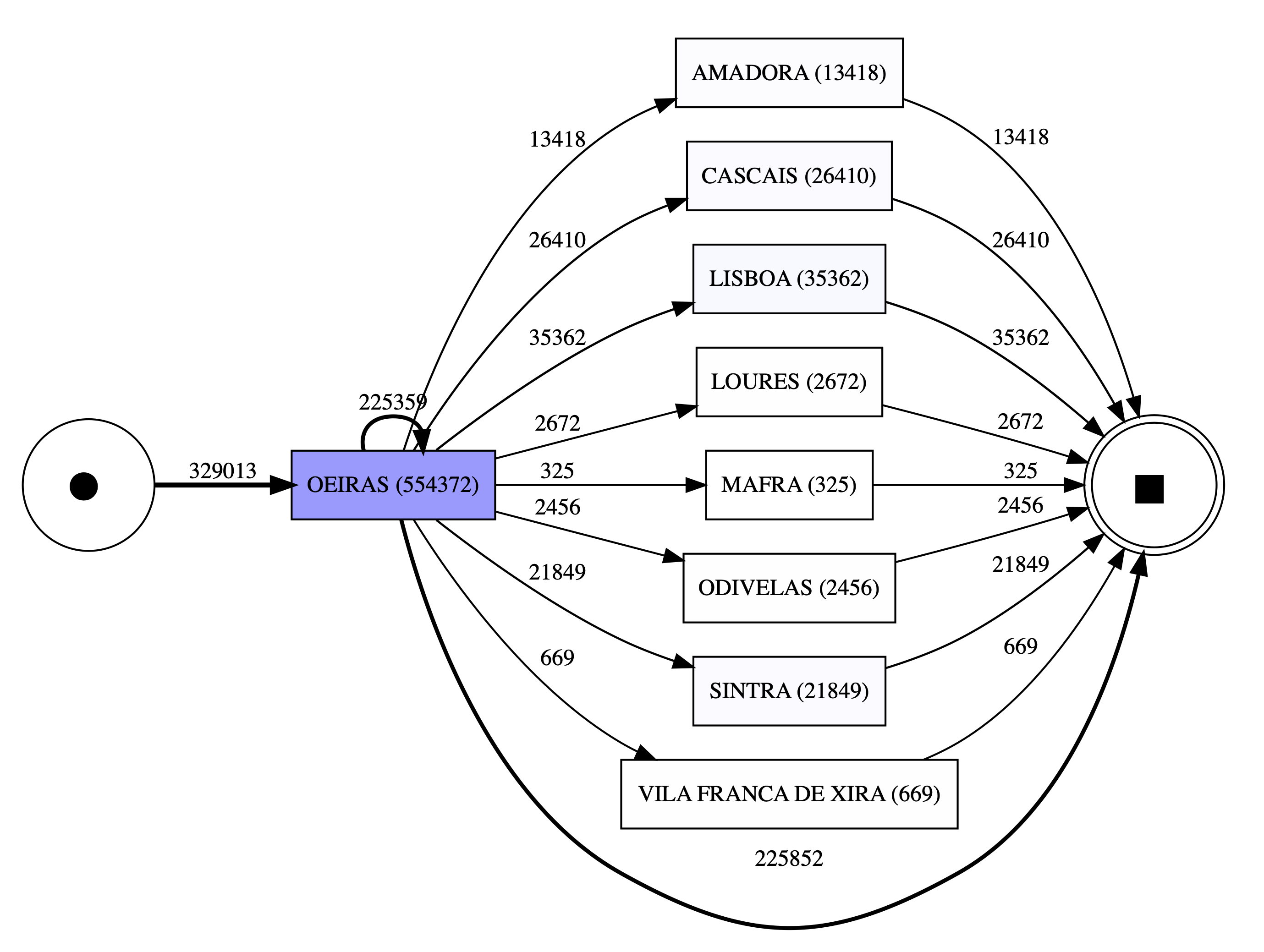}
    \caption{DFG by frequency}
    \label{fig:heu}
\end{subfigure}
\hfill
\begin{subfigure}[t]{0.5\textwidth}
    \centering
    \includegraphics[width=\linewidth]{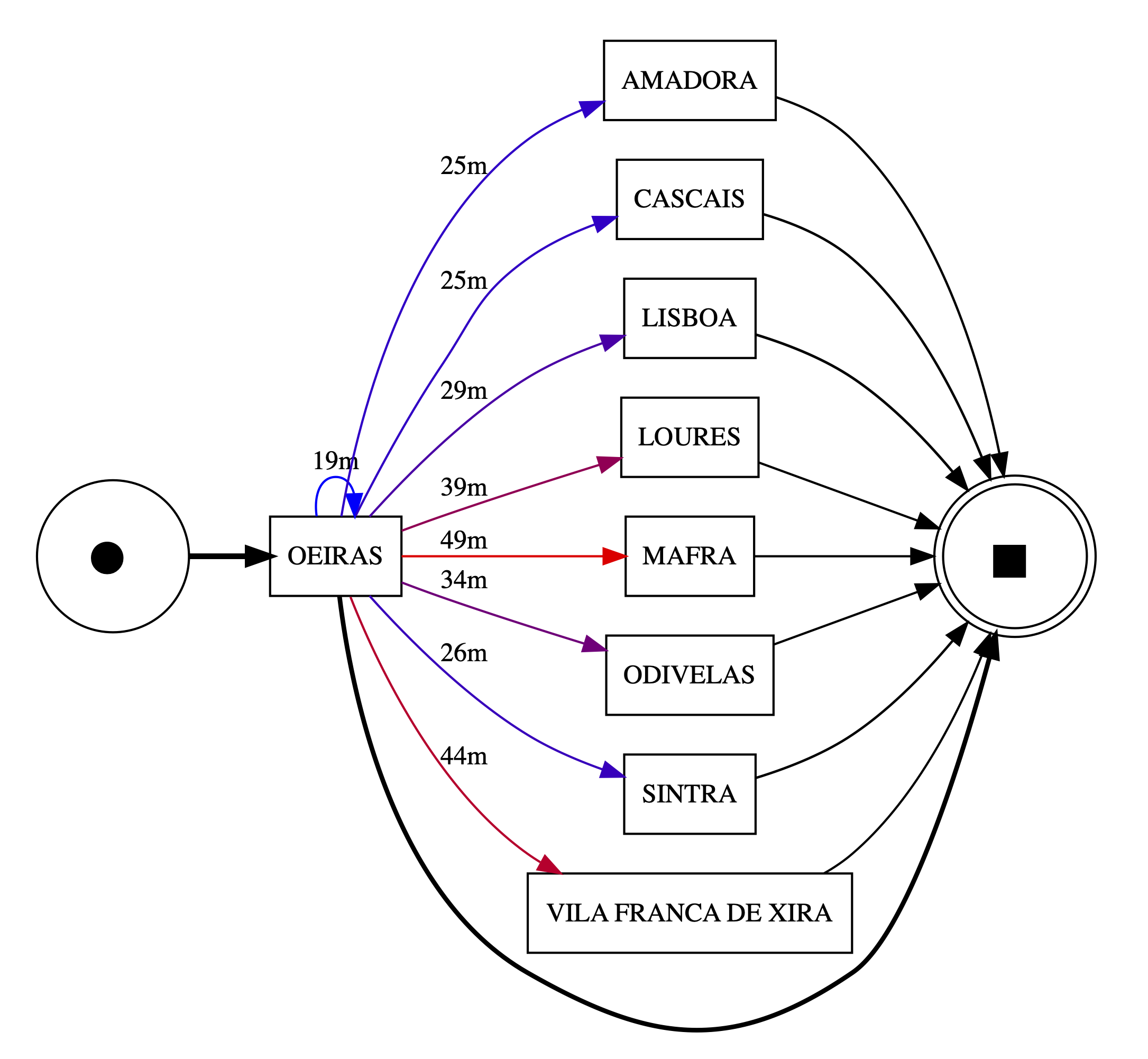}
    \caption{DFG with mean duration}
    \label{fig:heu_mun_oer}
\end{subfigure}
\caption{Municipality-level Directly-Follows Graphs: (a) by frequency and (b) with mean duration.}
\label{fig:heu_combined}
\end{figure}

The initial view (Fig.~\ref{fig:heu_top50fr}) shows only the top 20 variants, yet the full log exhibits 815 distinct variants, reflecting fine spatial/temporal granularity and high behavioral variability. To clarify the OD structure, we aggregated to the municipal level (Fig.~\ref{fig:heu}). In this abstraction, $554,372$ movements occur entirely within Oeiras (visible as a self-loop on node "Oeiras"), and $225,359$ trajectories terminate immediately after an intra-municipal move. The remaining flows are distributed primarily across Sintra, Lisbon, and Cascais. Contrary to the common assumption that suburban travel predominantly targets Lisbon, the inter-municipal pattern is more diverse. The mean travel time between Oeiras and Lisbon is 29 minutes, aligned with typical rail / car times along the coastal corridor.

\subsection{Object-centric DFGs}
Traditional event logs are centered on a single process instance (a Case ID). While effective in many domains, this one-case perspective cannot capture the concurrency and interdependencies of real processes~\cite{vanderaalstObjectCentricProcessMining2019}. In urban mobility, multiple objects—travelers, vehicles, and services—interact within infrastructure and contextual constraints, making case-centric logs particularly restrictive.

To better represent this complexity, we construct an object-centric event log (OCEL) and discover an object-centric directly-follows graph (OC-DFG) that links events through their participating objects~\cite{bertiOCPMAnalyzingObjectcentric2023,vanderaalstObjectCentricProcessMining2019}. Our current transport mode labels are derived from simple heuristics (e.g., speed, distance) and are therefore indicative rather than definitive: The goal is to demonstrate the applicability of OCPM, not to provide validated ground-truth labels.

We define fine-grained transport classes and use municipalities as activity labels to obtain a high-level view of regional flows. Within Oeiras, trips are plausibly dominated by bus and car. Because distinguishing these two modes using only kinematic features is difficult, the model tends to overlabel bus trips, whereas municipal survey statistics indicate that roughly half of trips are by car.

\begin{figure}[h!]
\centering
\includegraphics[width=6.5cm]{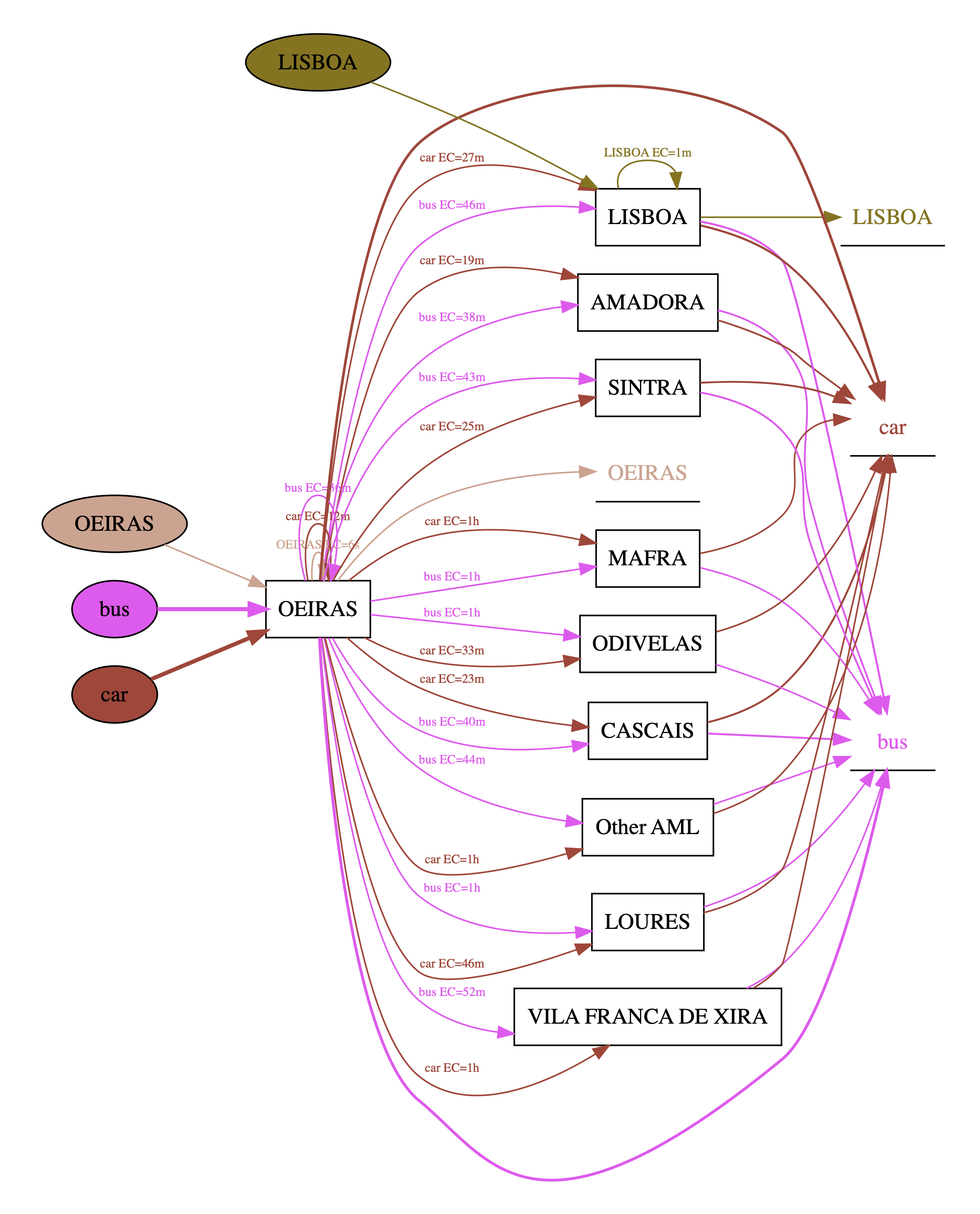}
\caption{OC-DFG — Bus vs Car (by duration)}
\label{fig:ocl-oeiras1}
\end{figure}

\begin{figure}[htb]
\centering
\includegraphics[width=6.5cm]{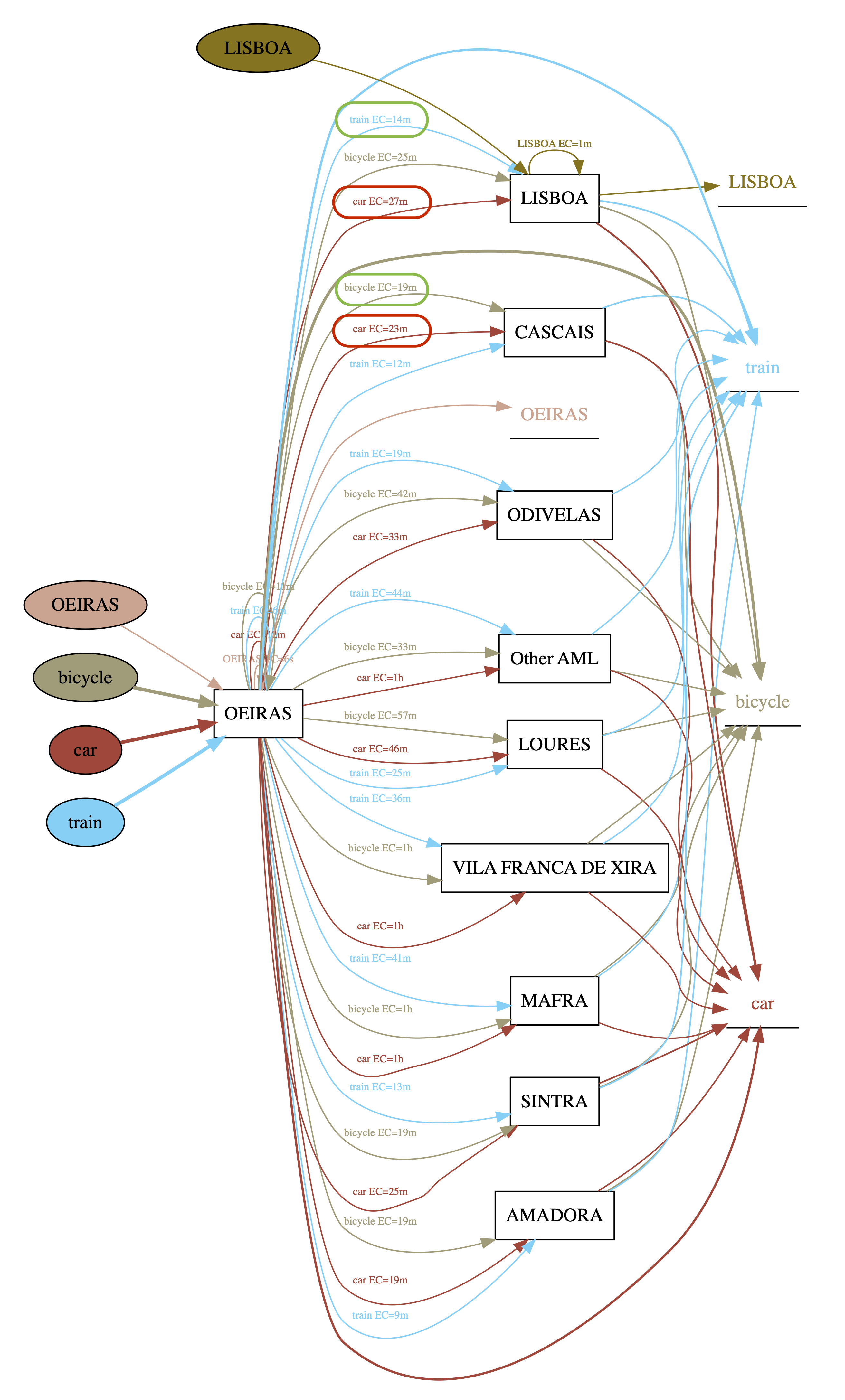}
\caption{OC-DFG — Train vs Car vs Bicycle (mean travel duration, min)}
\label{fig:ocl-oeiras2}
\end{figure}

Figure~\ref{fig:ocl-oeiras1} shows the median durations for bus trips (34 minutes) versus car (12 minutes) within Oeiras, consistent with indirect bus routes and the hilly topography of the area. Figure~\ref{fig:ocl-oeiras2} contrasts bicycle, car and train trips, illustrating the challenges of distinguishing electric bikes from motorized modes. The coastal rail line offers direct access to central Lisbon, complemented by an inland line in northern Oeiras. With more precise mode labeling, the OC model can help planners quantify duration differences across modes and identify opportunities to make non-car options more competitive.

\section{Comparison with survey of Oeiras}
\begin{figure}[h!]
    \centering
    \includegraphics[width=9.5cm]{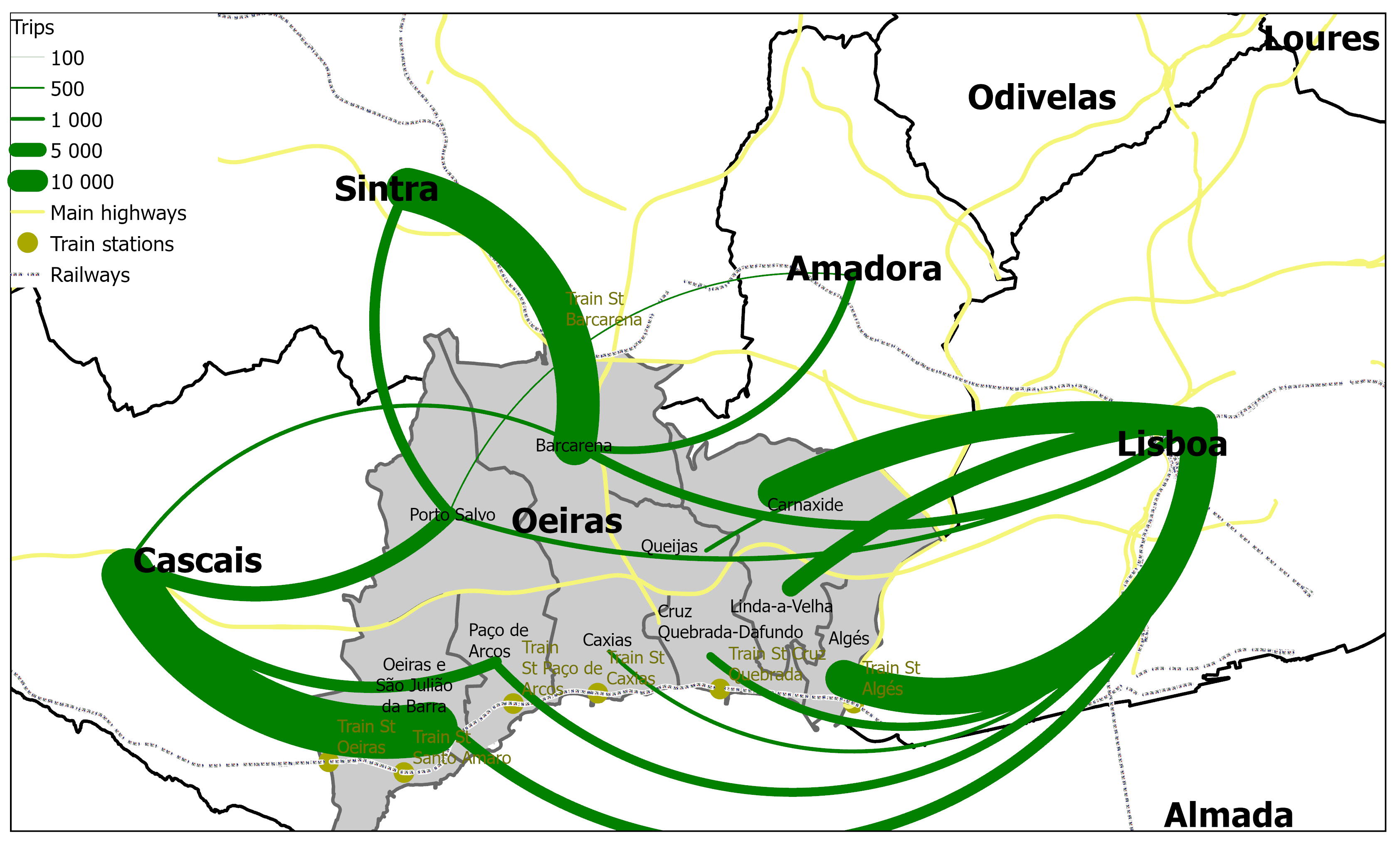}
    \caption{Destinations of trips originating in Oeiras within the Lisbon Metropolitan Area.}
    \label{fig:maptrips}
\end{figure}

%%%SHORT VERSION %%%%%
Figure \ref{fig:maptrips} displays the main OD pairs of trips obtained from CDRs made from Oeiras to other municipalities in the Lisbon Metropolitan Area (AML). To validate our results, we compared them with the 2019 Oeiras mobility survey\footnote{\url{https://www.oeiras.pt/documents/20124/2358115/Relatório+da+Fase+1+-+PMUS.pdf/b1120210-0ecd-d31e-e025-75903060f9eb?t=1685379606739}}, which interviewed 3,534 residents ($2.4$\% of the population). The survey reported 245,100 daily trips, of which $57.3$\% were intra-municipal, $26.9$\% to Lisbon, and $15.4$\% to other municipalities.
Compared to these data, our CDR-based approach tends to overestimate intra-municipal flows and trips between bordering municipalities (e.g., Barcarena–Sintra, Oeiras e São Julião–Cascais) while underestimating trips to Lisbon. This bias arises mainly from treating individual triplegs as full trips and from antenna coverage near municipal borders (e.g., train stations in Sintra or Cascais being assigned as destinations rather than intermediate legs).
To assess overall consistency, we applied linear regression between survey and CDR estimates, noting that minor discrepancies in parish definitions (2011 vs. current) prevented a one-to-one comparison. %The results are summarized in Table~\ref{tab:regression_stats}. 
The regression analysis demonstrates a very strong and statistically significant correlation between the number of trips obtained through the measurements and those reported in the survey ($R = 0.95, R^2 = 0.91, p < 0.001$). This indicates that the survey explains more 90\% of the variation in measured trips, confirming the reliability of the survey as a representation of actual mobility patterns. Although one sample shows a marked deviation (the trips from Oeiras to Lisbon), the overall agreement is robust, supporting the use of the survey data for further analysis.

\section{Conclusion \& Future Research }
Understanding human mobility is both challenging and essential for urban planning. In this study, we showed how Call Detail Records (CDRs) can be analyzed with process mining by representing trips either as traditional case-centric event logs (trip IDs) or as object-centric logs that link trips, modes, and locations. Our results demonstrate that mobility traces can indeed be structured as event logs, but current discovery techniques face difficulties with sparse data and data-preparation complexity.

Future work should focus on richer process models and tighter integration with machine learning. Large language models (LLMs) could assist in pattern discovery and interpretation from event sequences, while representing Object-Centric Event Logs (OCELs) as heterogeneous networks would extend beyond directly-follows graphs to capture multi-entity relations more naturally. Such representations open the door to tasks such as node classification, similarity search, and anomaly detection, and can be coupled with heterogeneous Graph Neural Networks (GNNs) for scalable analysis~\cite{dongHeterogeneousNetworkRepresentation2020,berti2023leveraginglargelanguagemodels,10899487}.

In general, our findings suggest that process mining provides valuable tools for analyzing multimodal travel behavior and can complement complex network approaches by adding a process-oriented perspective. For urban planners, this means new ways to explore travel sequences, compare mode-specific durations, and support policies that encourage efficient and sustainable mobility.

The preprocessing of CDRs\footnote{\url{https://github.com/kfilonchik/mobility4py}} and experiments using process mining\footnote{\url{https://github.com/kfilonchik/pm4mobility}} can be found in GitHub.

\section*{Acknowledgments}
The authors acknowledge the financial support provided by FCT Portugal under the project UIDB/04152/2020 – Centro de Investigação em Gestão de Informação (MagIC/NovaIMS) (https://doi.org/10.54499/UIDB/04152/2020) and Project Enhance - F -DUT-2022-0393.

% ---- OR ------- Uncomment this section to use bibtex
 \bibliographystyle{spmpsci} % We choose the "plain" reference style
 \bibliography{references_apa.bib} % Entries are in the refs.bib file
\end{document}